\documentclass[reprint,pra,amsmath,amssymb,aps,showpacs,preprintnumbers]{revtex4-1}
\usepackage{graphicx} 
\usepackage{datetime}
\usepackage{bm} 

\begin{document}
\title{Mode volume, energy transfer,  and spaser threshold in plasmonic systems with gain}
\author{Tigran V. Shahbazyan}
\affiliation{
Department of Physics, Jackson State University, Jackson, MS
39217 USA}


\begin{abstract} 
We present a unified approach to describe spasing in  plasmonic systems modeled by quantum emitters interacting with resonant plasmon mode. We show that spaser threshold implies detailed energy transfer balance between the gain and  plasmon mode and derive explicit spaser condition valid for arbitrary plasmonic systems. By defining carefully the plasmon mode volume relative to the gain region, we show that the spaser condition represents, in fact, the standard laser threshold condition extended to plasmonic systems with dispersive dielectric function. For extended gain region, the saturated mode volume depends solely on the system parameters that determine the lower bound of threshold population inversion.
\end{abstract}
\maketitle


%
\section{Introduction}
\label{sec:intro}
The prediction of plasmonic laser (spaser) \cite{bergman-prl03,stockman-natphot08,stockman-jo10} and its  experimental realization in various systems \cite{noginov-nature09,zhang-nature09,zheludev-oe09,zhang-natmat10,ning-prb12,gwo-science12,odom-natnano13,shalaev-nl13,gwo-nl14,zhang-natnano14,odom-natnano15} have been among the highlights in the rapidly developing field of plasmonics during the past decade \cite{stockman-review}. First observed in gold nanoparticles (NP) coated by dye-doped silics shells \cite{noginov-nature09}, spasing action was reported in hybrid plasmonic waveguides \cite{zhang-nature09}, semiconductor quantum dots on metal film \cite{zheludev-oe09,gwo-nl14},  plasmonic nanocavities and nanocavity arrays \cite{zhang-natmat10,ning-prb12,gwo-science12,odom-natnano13,zhang-natnano14,odom-natnano15},  and metallic NP and nanorods \cite{noginov-nature09,shalaev-nl13}, and more recently, carbon-based structures \cite{apalkov-light14,premaratne-acsnano14}. Small  spaser size well below the diffraction limit gives rise to wealth of applications  \cite{stockman-aop17}.

The spaser feedback mechanism is based on the near-field coupling between  resonant plasmon mode and gain medium, modeled here by an ensemble of  pumped two-level quantum emitters (QE) with  excitation frequency  tuned to the plasmon   frequency. The spaser threshold condition has been suggested as  \cite{bergman-prl03,stockman-natphot08,stockman-jo10} 
\begin{equation}
\label{laser-condition}
\frac{4\pi \mu^{2}\tau_{2} }{3\hbar }\frac{N_{21}}{\cal V }\,Q  \simeq 1,
\end{equation}
where $\mu$ and $\tau_{2}$ are the QE dipole matrix element and polarization relaxation time,  respectively, $N_{21}=N_{2}-N_{1}$ is the ensemble population inversion ($N_{2}$ and $N_{1}$ are, respectively, the number of excited and ground-state QEs), $Q$ is the  mode quality factor, and ${\cal V }$ is the mode volume.  Equation (\ref{laser-condition}) is similar to the standard laser condition \cite{haken} that determines the threshold value of $N_{21}$, but with  the cavity mode  quality factor and volume replaced by their plasmon counterparts in metal-dielectric system characterized by dispersive dielectric function $\varepsilon(\omega,\bm{r})$.  While the plasmon quality factor $Q$ is  well-defined   in terms of  the metal dielectric function $\varepsilon(\omega)=\varepsilon'(\omega)+i\varepsilon''(\omega)$, there is an active debate on mode volume definition  in plasmonic systems \cite{maier-oe06,polman-nl10,koenderink-ol10,hughes-ol12,russel-prb12,lalanne-prl13,hughes-acsphot14,bonod-prb15,muljarov-prb16,shegai-oe16,derex-jo16}.  Since  QEs are usually distributed outside the plasmonic structure, the standard expression for cavity mode volume, $\int dV \varepsilon(\bm{r}) |\textbf{E}(\bm{r})|^{2} /\text{max}[\varepsilon(\bm{r}) |\textbf{E}(\bm{r})|^{2}]$, where $\textbf{E}(\bm{r})$ is the mode electric field,  is ill-defined for open systems \cite{koenderink-ol10,hughes-ol12,hughes-acsphot14}. Furthermore, defining the  plasmon mode volume in terms of   field intensity at a specific point \cite{maier-oe06,lalanne-prl13} seems impractical due to very large local field variations near the metal surface caused by particulars of system geometry, for example, sharp edges or surface imperfections: strong  field fluctuations would grossly underestimate the mode volume that determines  spasing threshold for gain distributed in an \textit{extended} region. At the same time, while spasing was theoretically studied for several  specific systems \cite{wegener-oe08,stockman-jo10,klar-bjn13,li-prb13,lisyansky-oe13,bordo-pra13}, the general spaser condition was derived, in terms of  system parameters such as   permittivities and   optical constants, only for two-component systems \cite{stockman-prl11,stockman-review} without apparent relation to the  mode volume in Eq.~(\ref{laser-condition}).  Note that the actual spasing systems can be comprised of many components, so that the extension of the laser condition (\ref{laser-condition}) to plasmonics implies some procedure, valid for \textit{any} nanoplasmonic system, to determine the plasmon mode volume.

\begin{figure}[tb]
\begin{center}
\includegraphics[width=0.99\columnwidth]{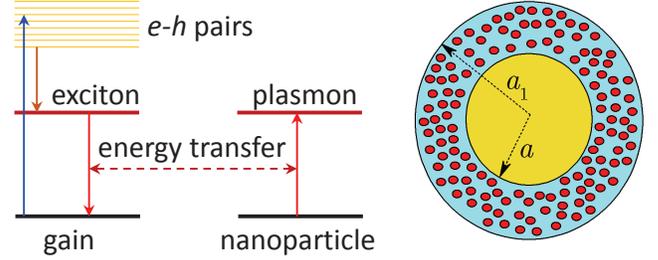}
\caption{\label{fig0} 
Spaser energy transfer diagram (left) and schematics for metal nanoparticle with dye-doped dielectric shell (right).}
\end{center}
\vspace{-8mm}
\end{figure}

On the other hand, the steady state spaser action implies detailed balance of  energy transfer (ET) processes between the QEs and the plasmon mode (see Fig.~\ref{fig0}). Whereas the energy flow between  individual QEs and plasmon  can go in either direction depending on the QE quantum state,  the \textit{net} gain-plasmon ET rate is determined by  population inversion $N_{21}$ and, importantly,   distribution of plasmon states in the gain region. Since  individual QE-plasmon ET rates are proportional to the plasmon local density of states (LDOS), which can vary in a wide range depending on  QEs' positions and system geometry \cite{shahbazyan-prl16}, the net   ET rate  is obtained by    \textit{averaging} the plasmon LDOS over the gain region. Therefore, the plasmon mode volume should relate, in terms of average  system characteristics, the  laser condition (\ref{laser-condition}) to the microscopic gain-plasmon ET picture. The goal of this paper is to establish such a relation.

%
%

First, we derive the general spaser condition for \textit{any} multicomponent nanoplasmonic system  in terms of individual ET rates  between QEs, constituting the gain, and resonant plasmon mode,  providing the feedback. Second, we introduce the plasmon mode volume $\mathcal{V}$ \textit{associated} with a region of volume $V_{0}$ by relating $\mathcal{V}$ to the plasmon LDOS, \textit{averaged} over that region, and establish that the spaser condition \textit{does} have the general  form (\ref{laser-condition}).  We then demonstrate, analytically and numerically, that a sufficiently extended region outside the plasmonic structure  can \textit{saturate} the plasmon mode volume, in which case $\mathcal{V}$ is \textit{independent} of the plasmon field distribution  and  determined solely by the system parameters,
\begin{equation}
\label{volume-mode-s}
\frac{{\cal V}}{V_{0}} 
 =Q\, \varepsilon_{d} \, \frac{\varepsilon''(\omega_{pl}) }{|\varepsilon' (\omega_{pl})|},
\end{equation}
where $\varepsilon_{d}$ is the gain region dielectric constant and  $\omega_{pl}$ is the plasmon frequency. With saturated mode volume (\ref{volume-mode-s}), the laser condition (\ref{laser-condition}) matches the spaser condition for two-component systems \cite{stockman-prl11,stockman-review} and, in fact, defines the \textit{lower bound} of threshold $N_{21}$. Finally, we demonstrate that, in realistic systems, the threshold  $N_{21}$ can  significantly exceed its minimal value.

\section{Spasing and gain-plasmon energy transfer balance }

We consider $N_{0}$ QEs  described by pumped two-level systems, located at positions $\textbf{r}_{j}$ near a plasmonic structure, with excitation energy $\hbar\omega_{21}=E_{2}-E_{1}$, where $E_{1}$ and $E_{2}$ are, respectively, the lower and upper level energies. Within the  density matrix approach, each QE is described by  polarization $\rho_{21}^{(j)}$ and occupation $n_{21}^{(j)}\equiv\rho_{22}^{(j)}-\rho_{11}^{(j)}$, so that  $N_{21}= N_{2}-N_{1}=\sum_{j}n_{21}^{(j)} $ is the ensemble population inversion. In the rotating wave approximation, the steady-state dynamics of QEs coupled to  alternating electric field $\bm{\mathcal{\cal E}}(\bm{r})e^{-i\omega t}$ is described by the Maxwell-Bloch equations 
\begin{align}
\label{maxwell-bloch}
& \left(\omega - \omega_{21} +i/\tau_{2}\right) \rho_{21}^{(j)} =\frac{\mu}{\hbar} \, n_{21}^{(j)} \,\bm{n}_{j}\!\cdot\! \bm{\mathcal{\cal E}}(\bm{r}_{j}),
\\
& n_{21}^{(j)} -\bar{n}_{21} =-  \frac{4\mu\tau_{1}}{\hbar} \,   \text{Im} \! \left[\rho_{21}^{(j)*} \,\bm{n}_{j}\!\cdot\! \bm{\mathcal{\cal E}}(\bm{r}_{j})\right],
\nonumber
\end{align}
where $\tau_2$ and  $\tau_1$ are the time constants characterizing polarization and population relaxation, $\mu$ and  $\bm{n}_{j}$ are, respectively, the QE dipole matrix element and orientation, and $\bar{n}_{21}$ is the average population inversion per QE due to the pump.  The  local field $\bm{\mathcal{\cal E}}(\bm{r}_{j})$ at the  QE position is generated by all QEs' dipole moments $\bm{p}_{j}=\mu \bm{n}_{j}\rho_{21}^{(j)}$ and, within semiclassical approach, has the form \cite{novotny-book} 
%
\begin{equation}
\bm{\mathcal{\cal E}}(\bm{r}_{j} ) =\frac{4\pi\omega^{2}}{c^{2}}\sum_{k} \bar{\textbf{G}}(\omega;\bm{r}_{j},\bm{r}_{k}) \!\cdot\! \bm{p}_{k},
\end{equation}
where $\bar{\textbf{G}}(\omega;\bm{r},\bm{r}')$ is the electromagnetic Green dyadic in the presence of metal nanostructure and $c$ is the speed of light. For nanoplasmonic systems, it is convenient to adopt  rescaled Green dyadic that has  direct near-field limit, $\bar{\textbf{D}}(\omega;\bm{r},\bm{r}')=-(4\pi\omega^{2}/c^{2})\bar{\textbf{G}}(\omega;\bm{r},\bm{r}')$. Upon eliminating the electric field, the system (\ref{maxwell-bloch}) takes the form
\begin{align}
\label{maxwell-bloch2}
&\Omega_{21}\bm{p}_{j}
+\frac{\mu^{2} }{\hbar}\, n_{21}^{(j)}\bm{n}_{j} \sum_{k}\bm{n}_{j} \!\cdot\!\bar{\textbf{D}}(\omega;\bm{r}_{j},\bm{r}_{k})\!\cdot\! \bm{p}_{k}
=0,
\nonumber\\
&\frac{\delta n_{21}^{j}}{\tau_{1}} -\frac{4}{\hbar}\,\text{Im}\sum_{k}\left[\bm{p}_{j}^{*}\!\cdot\! \bar{\textbf{D}}(\omega;\bm{r}_{j},\bm{r}_{k})\!\cdot\! \bm{p}_{k}\right]
= 0,
\end{align}
where we use shorthand notations $\Omega_{21}=\omega-\omega_{21}+i/\tau_{2}$  and $\delta n_{21}^{j}=n_{21}^{(j)}-\bar{n}_{21}$. The first equation in  system (\ref{maxwell-bloch2}), being homogeneous in $\bm{p}_{j}$, leads to the spaser threshold condition. Since the Green dyadic includes contributions from all electromagnetic modes, the spaser threshold in general case can only be determined numerically.  However, for QEs coupled to a \textit{resonant plasmon mode}, that is, for $\omega_{21}$ close to the mode frequency $\omega_{pl}$, the contribution from  off-resonance modes is relatively small \cite{pustovit-prb16,petrosyan-17} and, as we show below, the spaser condition can be obtained explicitly for any nanoplasmonic system.

\subsection{Gain coupling to a resonant plasmon mode}

For QE frequencies $\omega_{21}$   close to the  plasmon   frequency $\omega_{pl}$,  we can adopt the  single mode  approximation for the Green dyadic  \cite{shahbazyan-prl16}
\begin{equation}
\label{dyadic-mode}
\bar{\textbf{D}}(\omega;\bm{r},\bm{r}') = \frac{\omega_{pl}}{4 U}\frac{\textbf{E}(\bm{r})\otimes \textbf{E}^{*}(\bm{r}')}{\omega-\omega_{pl}+i/\tau_{pl}},
\end{equation}
where $\textbf{E}(\bm{r})$ is the slow envelope of plasmon field satisfying the Gauss  law $\bm{\nabla}\cdot \left [\varepsilon' (\omega_{pl},\bm{r}) \textbf{E}(\bm{r})\right ]=0$  and $1/\tau_{pl}$ is the plasmon   decay rate. In  nanoplasmonic systems, the  decay rate is dominated by the Ohmic losses  and  has the form
\begin{equation}
\label{plasmon-rate}
\dfrac{1}{\tau_{pl}}=\dfrac{W}{2U},
\end{equation}
where 
\begin{align}
\label{energy-mode}
U 
= \frac{1}{16\pi}\!\int \!  dV   \left |\textbf{E} (\bm{r})\right |^{2}  \partial \left [\omega_{pl}\varepsilon'(\omega_{pl},\bm{r})\right ]/\partial \omega_{pl} 
\end{align}
is the  mode stored energy, and 
\begin{equation}
\label{power-mode}
W=\frac{\omega_{pl}}{8\pi}\!\int \! dV  \left |\textbf{E}(\bm{r})\right |^{2}\varepsilon''(\omega_{pl},\bm{r})
\end{equation}
is the mode dissipated power \cite{landau}. The volume integration in $U$ and $W$ takes  place, in fact, only over the metallic regions with dispersive dielectric function.  For systems with a \textit{single} metallic region, one obtains the standard plasmon decay rate: 
%
\begin{equation}
\frac{1}{\tau_{pl}}=\frac{\varepsilon''(\omega_{pl})}{\partial \varepsilon'(\omega_{pl})/\partial \omega_{pl}}. 
\end{equation}
The Green dyadic (\ref{dyadic-mode})  is valid for a  well-defined plasmon mode ($\omega_{pl}\tau_{pl}\gg 1$) in any nanoplasmonic system, and its consistency is ensured by the optical theorem \cite{shahbazyan-prl16}.

With  the plasmon Green dyadic (\ref{dyadic-mode}),  the system (\ref{maxwell-bloch2}) takes the form
\begin{align}
\label{mb3}
&\Omega_{21}\bm{p}_{j} + \frac{ \mu^{2}}{\hbar} \frac{\omega_{pl}n_{21}^{(j)}}{4 U\Omega_{pl}}\, \bm{n}_{j}[\bm{n}_{j}\!\cdot\!\textbf{E}(\bm{r}_{j})]\sum_{k}\textbf{E}^{*}(\bm{r}_{k})\!\cdot\! \bm{p}_{k}
=0,
\nonumber\\
&\frac{\delta n_{21}^{j}}{\tau_{1}} 
- \text{Im} \biggl [\frac{\omega_{pl}}
{\hbar U\Omega_{pl}}
\, [\bm{p}_{j}^{*}\!\cdot\!\textbf{E}(\bm{r}_{j})]
\sum_{k}\textbf{E}^{*}(\bm{r}_{k})\!\cdot\! \bm{p}_{k}\biggr ]
=0,
\end{align}
where  $\Omega_{pl}=\omega-\omega_{pl}+i/\tau_{pl}$. Multiplying the first equation by $\textbf{E}^{*}(\bm{r}_{j})$ and summing up over $j$, we obtain the  spaser condition
\begin{align}
\label{d}
\Omega_{21}\Omega_{pl}+ \frac{\mu^{2}}{\hbar} 
\frac{\omega_{pl}}{4 U} 
\sum_{j}n_{21}^{(j)}  
 \left |\bm{n}_{j}\!\cdot\!\textbf{E}(\bm{r}_{j})\right |^{2}=0.
\end{align}
The second term in Eq.~(\ref{d}) describes coherent coupling between the QE ensemble and plasmon mode. Below we show that spasing implies  detailed ET balance between the gain and plasmon mode.

\subsection{Energy transfer and spaser condition}

Let us now introduce, in the standard manner, the individual QE-plasmon ET rate as \cite{shahbazyan-prl16}
%
\begin{equation}
\label{rate-mode}
\frac{1}{\tau} 
=-\frac{\mu^{2}}{\hbar}\text{Im} \! \left [
\bm{n}\!\cdot\!\bar{\textbf{D}}(\omega_{pl};\bm{r},\bm{r})\!\cdot\!\bm{n}\right ]
=\!\frac{4\pi\mu^{2}}{\hbar}
\frac{|\bm{n}\!\cdot\!\textbf{E}(\bm{r})|^{2}}{\int \!dV \varepsilon''|\textbf{E}|^{2} },
\end{equation}
where we used Eqs.~(\ref{dyadic-mode}) and  (\ref{plasmon-rate}), and implied $\varepsilon\equiv \varepsilon(\omega_{pl},\bm{r})$ under the integral. The condition (\ref{d}) can  be recast as
\begin{equation}
\label{cond-gen}
\left (\omega-\omega_{21} +\frac{i}{\tau_{2}}\right ) \left (\omega-\omega_{pl}+\frac{i}{\tau_{pl}}\right )+\frac{1}{\tau_{ g}\tau_{pl}} =0,
\end{equation}
where we introduced  \textit{net} gain-plasmon ET rate, 
\begin{equation}
\label{et-net}
\frac{1}{\tau_{g}}
=\sum_{j}  \frac{n_{21}^{(j)}}{\tau_{j}}
= \frac{4\pi\mu^{2}}{\hbar}\sum_{j}  n_{21}^{(j)} \,
\frac{|\bm{n}_{j}\!\cdot\!\textbf{E}(\bm{r}_{j})|^{2}}{\int \!dV \varepsilon''|\textbf{E}|^{2} },
\end{equation}
which represents the sum of individual QE-plasmon ET rates $1/\tau_{j}$, given by Eq.~(\ref{rate-mode}), weighed by  QE occupation numbers. Since $n_{21}^{(j)}$ is positive or negative for   QE  in the  excited or ground state, respectively, the direction of energy flow between the QE and the plasmon mode depends on  the QE  quantum state. Note that the main contribution to $1/\tau_{g}$ comes from  the regions with  large  plasmon LDOS, that is,  high  QE-plasmon ET rates (\ref{rate-mode}). The imaginary part of Eq.~(\ref{cond-gen})  yields the spaser frequency \cite{bergman-prl03}
\begin{equation}
\label{resonant}
\omega_{s}=\frac{\omega_{pl}\tau_{pl}+\omega_{21}\tau_{2}}{\tau_{pl}+\tau_{2}},
\end{equation}
while its real part, with the above $\omega_{s}$, leads to  
\begin{equation}
\label{spaser-cond}
\frac{1}{\tau_{g}\tau_{pl}}=\frac{1}{\tau_{2}\tau_{pl}} + \frac{\left (\omega_{pl}-\omega_{21}\right)^{2}\tau_{2}\tau_{pl}}{(\tau_{pl}+\tau_{2})^{2}}.
\end{equation}
In  the case when the QE  and plasmon spectral bands overlap well,  that is.,  $|\omega_{pl}-\omega_{21}|\tau_{pl}\ll 1 $ or $|\omega_{pl}-\omega_{21}|\tau_{2}\ll 1 $ depending on relative magnitude of the respective bandwidths $1/\tau_{2}$ and $1/\tau_{pl}$, the last term in Eq.~(\ref{spaser-cond}) can be disregarded,  and we arrive at the spaser condition in the form   $1/\tau_{g}=1/\tau_{2}$, or
\begin{equation}
\label{condition}
\sum_{j}  \frac{n_{21}^{(j)}}{\tau_{j}}
=\frac{1}{\tau_{2}}.
\end{equation}
Equation (\ref{condition}) implies that   spaser threshold is reached when   \textit{energy transfer balance} between gain and  plasmon mode is established.

\subsection{System geometry and QE-plasmon ET rate }

%
\begin{figure}[tb]
\begin{center}
\includegraphics[width=0.99\columnwidth]{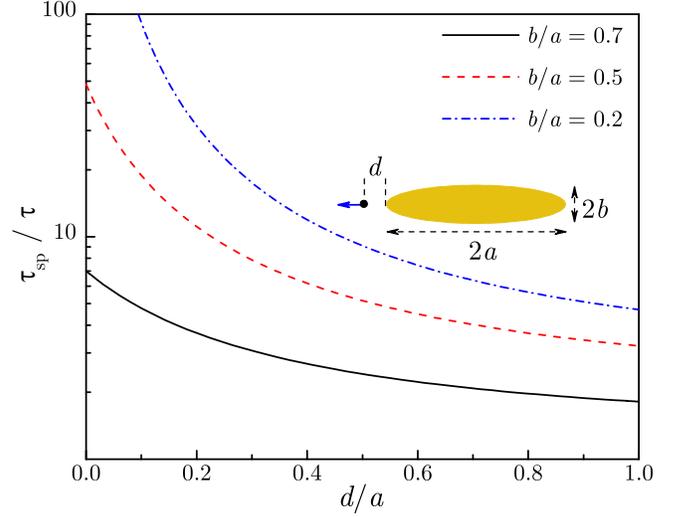}
\caption{\label{fig1} 
QE-plasmon ET rate (\ref{rate-mode}) for a QE near a spheroidal particle with aspect ratio $b/a$ normalized by that for spherical particle with radius $a$. }
\end{center}
\vspace{-8mm}
\end{figure}
%

Individual ET rates in the spaser condition (\ref{condition}) can vary in a wide range depending on the QE position and system geometry. In Fig.~\ref{fig1}, we show the ET rate (\ref{rate-mode}) for a QE located at  distance $d$ from  a tip of   gold nanorod, modeled here by  prolate spheroid with semiaxes $a$ and $b$  (see Appendix). In all numerical calculations, we use the experimental dielectric function for gold \cite{christy}. To highlight the role of system geometry, the ET rate $1/\tau$ for nanorod is normalized by the ET rate $1/\tau_{sp}$ for  sphere of radius $a$. The latter ET rate has the form  
\begin{equation}
\frac{1}{\tau_{sp}}=\frac{12\mu^{2}}{\hbar \varepsilon''(\omega_{sp})}\frac{a^{3}}{(a+d)^{6}},
\end{equation}
and experiences a sharp decrease for $d>a$. With changing nanoparticle shape,  three degenerate dipole modes of a sphere split into longitudinal and two transverse modes. The latter move up in energy to get damped by  interband transitions in gold with their onset just above the plasmon energy in spherical particles, while the longitudinal mode moves down in energy away from the transitions onset, thereby gaining in the oscillator strength \cite{mulvaney-prl02}. This sharpening of plasmon resonance together with condensation of plasmon states near the tips (lightning rod effect)   results in up to 100-fold rate increase with reducing $b/a$ ratio, as shown in Fig.~\ref{fig1}, indicating that spasing is dominated by QEs  located in the large plasmon LDOS regions. Large variations of $1/\tau$ magnitude imply that the plasmon mode volume, which characterizes spatial extent of the gain region with sufficiently strong QE-plasmon coupling, is determined by the \textit{average} plasmon LDOS  in that region, as we show in the next section.

\section{Plasmon mode volume and spaser threshold}

\subsection{General spaser condition}

The form (\ref{condition}) of spaser condition reveals the microscopic origin of spaser action as the result of cooperative ET between  gain  and resonant plasmon mode, with each QE contribution depending on its position and quantum state. Below we assume that QEs  are distributed within some region of volume $V_{0}$ and that  population inversion distribution follows, on average, that of QEs. After averaging   over   QEs' dipole orientations, the gain-plasmon  ET rate (\ref{et-net}) takes the form
\begin{equation}
\label{et-net2}
\frac{1}{\tau_{g}}=\frac{ 4\pi\mu^{2}}{3\hbar }   \frac{ \int \! dV_{0} n_{21} (\bm{r})|\textbf{E}(\bm{r}) |^{2}}{\int\! dV \varepsilon''(\omega_{pl},\bm{r})| \textbf{E}(\bm{r})|^{2}},
\end{equation}
where $n_{21} (\bm{r})$ is  population inversion \textit{density}, yielding the spaser threshold condition 
\begin{equation}
\label{condition-spaser}
\frac{ 4\pi\mu^{2}\tau_{2}}{3\hbar }   \frac{ \int \! dV_{0} n_{21} (\bm{r})|\textbf{E}(\bm{r}) |^{2}}{\int\! dV \varepsilon''(\omega_{pl},\bm{r})| \textbf{E}(\bm{r})|^{2}} =1,
\end{equation}
which is valid for \textit{any} multicomponent system supporting a well-defined surface plasmon. In the case of \textit{uniform} gain distribution,  $n_{21}=N_{21}/V_{0}$, and a \textit{single} metallic component with volume $V_{m}$ [e.g., a metal particle with dye-doped dielectric shell (see Fig.~\ref{fig0})], the threshold condition (\ref{condition-spaser}) takes the form 
\begin{equation}
\label{condition1}
\frac{ 4\pi\mu^{2}\tau_{2}}{3\hbar }\, \frac{n_{21}}{\varepsilon''(\omega_{pl})}\, \frac{\int\! dV_{0}  |\textbf{E}|^{2}}{\int\! dV_{m} |\textbf{E}|^{2}} =1.
\end{equation}
The threshold  value of $n_{21}$  is determined by   ratio of plasmon field integral intensities in the gain and metal regions. Evidently, the spaser threshold \textit{does} depend on the gain region size and shape, which prompts us to revisit the mode volume definition for plasmonic systems in order to ensure its consistency with the general laser condition (\ref{laser-condition}).

\subsection{Plasmon LDOS   and associated mode volume}

Here, we show that the mode volume in plasmonic systems  can be accurately defined in terms of  plasmon LDOS.  The LDOS of a single plasmon mode is related  to the  plasmon Green dyadic (\ref{dyadic-mode}) as $\rho (\omega,\bm{r})
=-(2\pi^{2} \omega_{pl})^{-1} \, \text{Im}\,\text{Tr}\, \bar{\textbf{D}}(\omega;\bm{r},\bm{r})$, and  has the Lorentzian form \cite{shahbazyan-prl16},
\begin{equation}
\label{ldos-mode}
\rho (\omega,\bm{r})
=\frac{\tau_{pl}}{8\pi^{2}  U} \frac{ \left |\textbf{E}(\bm{r})\right |^{2}}{(\omega-\omega_{pl})^{2}\tau_{pl}^{2}+1},
\end{equation}
where $U$ is given by Eq.~(\ref{energy-mode}).  The plasmon LDOS (\ref{ldos-mode}) characterizes the distribution of plasmon states in the unit volume and frequency interval. Consequently, its frequency integral, $\rho (\bm{r})=\!\int\! d\omega \rho (\omega,\bm{r})$, represents the  \textit{plasmon mode density} that describes the plasmon states' spatial distribution:
\begin{equation}
\label{density}
\rho  (\bm{r})=\frac{\left |\textbf{E} (\bm{r})\right |^{2}}{8\pi U }
= \frac{2\left |\textbf{E} (\bm{r})\right |^{2}}{\int \! dV [\partial (\omega_{pl}\varepsilon')/\partial \omega_{pl}]|\textbf{E}|^{2}}.
\end{equation}
Introducing the mode quality factor $Q=\omega_{pl} U/W$, the mode density can be written as 
\begin{equation}
\label{density1}
\rho  (\bm{r}) 
= \frac{1}{Q}\frac{\left |\textbf{E} (\bm{r})\right |^{2}}{\int \! dV  \varepsilon'' |\textbf{E}|^{2}}.
\end{equation}
Note that, in terms of $\rho  (\bm{r})$, the gain-plasmon ET rate (\ref{et-net2}) takes the form
\begin{equation}
\label{et-net3}
\frac{1}{\tau_{g}}=\frac{ 4\pi\mu^{2}}{3\hbar }\, Q   
\int \! dV_{0} n_{21} (\bm{r}) \rho(\bm{r}),
\end{equation}
implying that the largest contribution to $1/\tau_{g}$ comes from QEs located in the regions with high plasmon density.

We now relate the plasmon mode volume ${\cal V} $ \textit{associated} with region $V_{0}$ to the \textit{average}  mode density in that region:
\begin{equation}
\label{volume-mode}
\frac{1}{\mathcal{V}}
= \frac{1}{V_{0}}\!\int\! dV_{0}   \rho(\bm{r}) 
= \frac{1}{V_{0}}\frac{2\int \! dV_{0}\left |\textbf{E}\right |^{2}}{\int \! dV [\partial ( \omega_{pl}\varepsilon')/\partial  \omega_{pl}]|\textbf{E}|^{2}},
\end{equation}
or, equivalently,
\begin{equation}
\label{volume-mode2}
\frac{\mathcal{V}}{V_{0}} 
 =Q \,  \frac{ \int \! dV  \varepsilon'' |\textbf{E}|^{2}}{\int\! dV_{0} |\textbf{E}|^{2}}.
\end{equation}
The expressions (\ref{volume-mode}) or (\ref{volume-mode2}) are valid for nanoplasmonic systems of any size and shape and with any number of metallic and dielectric components.

It is straightforward to check that, for uniform gain distribution with $n_{21}=N_{21}/V_{0}$, the spaser threshold condition (\ref{condition-spaser}) \textit{coincides} with the laser condition (\ref{laser-condition}) with associated mode volume $\mathcal{V}$ given by Eq.~(\ref{volume-mode2}). Equivalently, for uniform gain distribution, the gain-plasmon ET rate (\ref{et-net3}) takes the form
\begin{equation}
\label{et-net4}
\frac{1}{\tau_{g}}=\frac{ 4\pi\mu^{2}}{3\hbar }\, \frac{N_{21}}{\mathcal{V}}\, Q,
\end{equation}
and the laser condition (\ref{laser-condition}) follows from the ET balance condition $1/\tau_{g}=1/\tau_{2}$. 

For systems with  single metal  component, the plasmon mode volume takes the form [compare to Eq.~(\ref{condition1})]
\begin{equation}
\label{volume-mode3}
\frac{\mathcal{V}}{V_{0}} 
=\frac{\omega_{pl}}{2}   \frac{\partial \varepsilon'(\omega_{pl})}{\partial \omega_{pl}} 
 \frac{ \int \! dV_{m}  |\textbf{E}|^{2}}{\int\! dV_{0} |\textbf{E}|^{2}}
 =Q \,  \varepsilon''(\omega_{pl})\,\frac{ \int \! dV_{m}  |\textbf{E}|^{2}}{\int\! dV_{0} |\textbf{E}|^{2}},
\end{equation}
where the plasmon quality factor has the form 
%
\begin{equation}
Q  = \omega_{pl}\, \frac{\partial \varepsilon'(\omega_{pl}) /\partial \omega_{pl}}{2\varepsilon'' (\omega_{pl}) }=\frac{\omega_{pl}\tau_{pl}}{2}.
\end{equation}
Note that, for a well-defined plasmon  with $Q\gg 1$, the plasmon mode volume is \textit{independent}  of Ohmic losses in metal.

\subsection{Mode volume saturation and lower bound of  spaser threshold}

Since the QE-plasmon ET rate rapidly falls outside the plasmonic structure (see Fig.~\ref{fig1}), spasing is dominated by QEs located sufficiently close to the metal surface. In the case when a metal nanostructure of volume $V_{m}$ is surrounded by an extended gain region $V_{0}$, so that the plasmon LDOS spillover beyond $V_{0}$ is negligible,  the  plasmon mode volume is \textit{saturated} by the gain, leading to \textit{constant} value of ${\cal V}/V_{0}$ that is \textit{independent} of the plasmon field distribution. To demonstrate this point, we note  that, in the quasistatic approximation,  the integrals in Eq.~(\ref{volume-mode3})  reduce  to surface terms, 
\begin{align}
\label{gauss}
&\int \! dV_{0}  |\textbf{E}|^{2}= \int \! dS\, \Phi ^{*} \nabla_{n} \Phi  + \!\int\! dS_{1} \Phi^{*} \nabla_{n} \Phi ,
\nonumber\\
&\int \! dV_{m}  |\textbf{E} |^{2}=\int \! dS\, \Phi ^{*} \nabla_{n} \Phi ,
\end{align}
where $S$ is the \textit{common} interface separating the metal and dielectric regions, $S_{1}$ is the outer  boundary of the dielectric region, $\Phi$ is the potential related to the plasmon field as   $\textbf{E}=-\bm{\nabla}\Phi$, and $\nabla_{n} \Phi$ is its normal derivative relative to the interface.  The potentials in the first and second equations of system (\ref{gauss}) are taken, respectively, at the dielectric and metal sides of the interface $S$. Since the plasmon fields rapidly fall  away from the metal,  the  contribution from the outer interface $S_{1}$ can be neglected for extended dielectric regions (see below). Then, using the standard boundary conditions at the common  interface $S$, we obtain from Eqs.~(\ref{volume-mode3})  and (\ref{gauss})   the \textit{saturated} mode volume:
\begin{equation}
\label{volume-mode4}
\frac{\mathcal{V}}{V_{0}} 
=\frac{\omega_{pl} \varepsilon_{d}}{2|\varepsilon'(\omega_{pl})| }   \frac{\partial \varepsilon'(\omega_{pl}) }{\partial \omega_{pl} } 
 =Q\, \varepsilon_{d} \, \frac{\varepsilon''(\omega_{pl}) }{|\varepsilon' (\omega_{pl})|}.
\end{equation}
Remarkably, the saturated mode volume depends on system geometry  only via the plasmon frequency $\omega_{pl}$ in the metal dielectric function. Combining Eqs.~(\ref{volume-mode4}) and (\ref{laser-condition}), we arrive at the spaser condition for saturated case,
\begin{equation}
\label{condition2}
\frac{ 4\pi\mu^{2}\tau_{2}}{3\hbar \varepsilon_{d}} \frac{|\varepsilon'(\omega_{pl})|}{\varepsilon''(\omega_{pl})} \,  n_{21} =1,
\end{equation}
which matches the spaser condition obtained previously for two-component systems, that is, with gain region extended to infinity \cite{stockman-review,stockman-prl11}. We stress  that the condition (\ref{condition2}) provides the \textit{lower bound} for  threshold value of $n_{21}$, while in real systems, where plasmon field distribution can extend beyond the gain region, the threshold   can be significantly higher, as we illustrate below.

%
\begin{figure}[tb]
\begin{center}
\includegraphics[width=0.99\columnwidth]{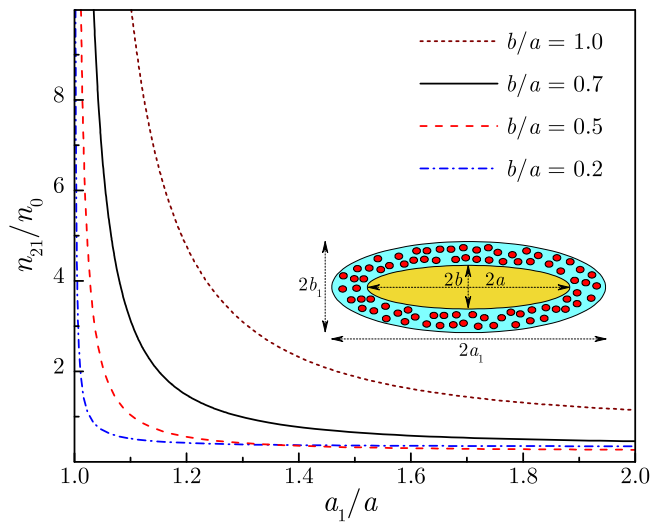}
\caption{\label{fig2} 
Calculated threshold $n_{21}$ for gold nanorods with dye-doped silica shell is shown with increasing gain region size. Rapid plasmon mode volume saturation for  small $b/a$ is due to condensation of plasmon states near the  nanorod tips. }
\end{center}
\vspace{-8mm}
\end{figure}
%
 
In Fig.~(\ref{fig2}), we show the change of threshold $n_{21}$ with expanding gain region in nanorod-based spaser modeled by composite spheroidal particle with gold core and QE-doped silica shell.  Calculations were performed using  Eq.~(\ref{condition1}) for confocal spheroids (see Appendix for details), and  QE   frequency $\omega_{21}$ was tuned to  resonance with  longitudinal dipole  mode  frequency $\omega_{pl}$. Note that the gain optical constants enter the spaser condition (\ref{condition1}) through a single parameter
\begin{equation}
n_{0}=  \frac{3\hbar}{4\pi\mu^{2}\tau_{2}},
\end{equation}
which represents characteristic gain concentration and sets the overall scale of threshold $n_{21}$ for a  specific gain medium. The ratio $n_{21}/n_{0}$, plotted in Fig.~\ref{fig2},  depends only on  plasmonic system parameters  and, with expanding gain region,  decreases prior reaching plateau corresponding to the saturated mode volume regime described by Eq.~(\ref{condition2}).  Note that, in nanorods, the rapid mode volume saturation seen in Fig.~\ref{fig2}, as compared to spherical particles, is caused by   condensation of plasmon states near the tips (lightning rod effect), leading to the much larger plasmon LDOS  and, correspondingly, the QE-plasmon ET rate (see Fig.~\ref{fig1}).

\section{Conclusions}

In summary, we have  developed a unified approach to spasing in a system of pumped quantum emitters interacting with a plasmonic structure of arbitrary shape in terms of energy transfer processes within the system. The threshold value of population inversion is determined from the condition of detailed energy transfer balance between quantum emitters, constituting the gain, and resonant plasmon mode, providing the feedback. We have shown that, in plasmonic systems, the mode volume should be defined relative to a finite region, rather than  to a point of maximal field intensity, by averaging the plasmon local density of states over that region. We demonstrated that, in terms of  plasmon mode volume, the spaser condition has the standard form of the laser threshold condition, thus,  extending the latter to plasmonic systems with dispersive dielectric function. We have also shown that, for extended gain region, the saturated plasmon mode volume is determined solely by the system permittivities, which define the lower bound of threshold  population inversion.

\acknowledgments
This work was supported in part by NSF grants No. DMR-1610427 and No. HRD-1547754.

\appendix

\section{Calculation of QE-plasmon ET rate for spheroidal nanoparticle}

The ET rate between a plasmon mode in metal nanoparticle with frequency $\omega_{pl}$ and a QE  located at the point $\bm{r}$ distanced by $d$ from the metal surface and  polarized along the \textit{normal} $\bm{n}$ to the  surface is given by
\begin{equation}
\label{ldos-pr-a}
\frac{1}{\tau}
=
\frac{4\pi\mu^{2}}{\hbar}
\frac{|\bm{n} \!\cdot\!\textbf{E}(\bm{r} )|^{2}}{\int \!dV \varepsilon''|\textbf{E}|^{2} }
=
\frac{4\pi\mu^{2}}{\hbar\varepsilon''(\omega_{pl})}
\frac{|{\nabla}_{n}\Phi(\bm{r} )|^{2}}{\int \!dS  \Phi^{*} {\nabla}_{n}\Phi},
\end{equation}
where $\nabla_{n}=\bm{n}\!\cdot\! \bm{\nabla}$ stands for the normal derivative, and real part of the denominator is implied.

Consider  a QE  at distance $d$ from the tip of a spheroidal particle with semiaxis $a$ along the symmetry axis and semiaxis  $b$ in the symmetry plane ($a>b$). The potentials have the form $\Phi\propto R_{lm}(\xi)Y_{lm}(\eta,\zeta)$, where $\xi$ is the radial (normal) coordinate and the pair $(\eta,\zeta)$ parametrizes the surface ($Y_{lm}$ are spherical harmonics). The surface area element is $dS=h_{\eta}h_{\zeta} d\eta d\zeta$, and normal derivative is $\nabla_{n}=h_{\xi}^{-1}(\partial/\partial\xi)$, where $h_{i}$ are the scale factors ($i=\xi,\eta,\zeta$)  given by 
\begin{align}
&h_{\xi}=f\sqrt{\frac{\xi^{2}-\eta^{2}}{\xi^{2}-1}},
~~
h_{\eta}=f\sqrt{\frac{\xi^{2}-\eta^{2}}{1-\eta^{2}}},
\nonumber\\
&h_{\zeta}=f\sqrt{(\xi^{2}-1)(1-\eta^{2})},
\end{align}
$f =\sqrt{a^{2}-b ^{2}}$ is half distance between the foci, and spheroid surface corresponds to $\xi_{1}=a/f$. 

For QE located at point $z=f\xi$ on the $z$-axis ($\eta=1$) outside the spheroid, the radial potentials for  dipole longitudinal plasmon mode $(lm)=(10)$ have the form  $R(\xi)=P_{1}(\xi)$ for $\xi<\xi_{1}$ and $R(\xi)=Q_{1}(\xi)P_{1}(\xi_{1})/Q_{1}(\xi_{1})$ for $\xi>\xi_{1}$,  where $P_{l}$ and $Q_{l}$ are the  Legendre functions of first and second kind, given by 
\begin{align}
&P_{1}(\xi)=\xi,
~~
Q_{1}(\xi)=\frac{\xi}{2}\ln\left (\frac{\xi+1}{\xi-1}\right )-1,
\nonumber\\
&Q'_{1}(\xi)=\frac{1}{2}\ln\left (\frac{\xi+1}{\xi-1}\right )-\frac{\xi}{\xi^{2}-1}.
\end{align}
Using $h_{\xi}=f$  along the $z$-axis, the ET rate equals
\begin{equation}
\label{rate-normal2-a}
\frac{1}{\tau}=
\frac{3\mu^{2}}{\hbar\varepsilon''(\omega_{pl})}
\frac{R^{\prime 2}(\xi)}{f^{3}\xi_{1}(\xi_{1}^{2}-1)}
 = 
\frac{3\mu^{2}}{\hbar ab^{2}\varepsilon''(\omega_{pl})}
\left [\frac{Q'_{1}(\xi)\xi_{1}}{Q_{1}(\xi_{1})}\right ]^{2},
\end{equation}
with $\xi=(a+d)/f$, where  the plasmon frequency $\omega_{pl}$ determined by the boundary condition $\varepsilon'(\omega_{pl})=\varepsilon_{d} Q'_{1}(\xi_{1})$. In the limit of spherical particle of radius $a$, that is, $f\rightarrow 0$ and $\xi \rightarrow \infty$ as $b\rightarrow a$, we have $Q(\xi)\approx 1/3\xi^{2}$, yielding 
\begin{equation}
\label{rate-sphere-a}
\frac{1}{\tau_{sp}}=
\frac{12\mu^{2}}{\hbar \varepsilon''(\omega_{sp})}
 \frac{a^{3}}{(a+d)^{6}},
\end{equation}
where $\omega_{sp}$ is surface plasmon resonance frequency for a sphere determined by $\varepsilon'(\omega_{sp})+2\varepsilon_{d}=0$. The normalized ET rate $\tau_{sp}/\tau$ has the form
\begin{equation}
\label{rate-normalized-a}
\frac{\tau_{sp}}{\tau}
=
\frac{a^{2}}{4b^{2}}
\left (1+\frac{d}{a}\right )^{6}
\frac{\varepsilon''(\omega_{sp})}{\varepsilon''(\omega_{pl})}
\left [\frac{Q'_{1}(\xi)\xi_{1}}{Q_{1}(\xi_{1})}\right ]^{2},
\end{equation}
with $\xi=\xi_{1}+d/f$.

\section{Calculation of population inversion density in spheroidal core-shell nanoparticle}

We consider a core-shell nanoparticle with dielectric functions $\varepsilon_{c}$, $\varepsilon_{s}$, and $\varepsilon_{d}$ in the core, shell, and outside dielectric, respectively, with inner and outer interface $S_{1}$ and $S_{2}$. The integrals over  core ans shell regions in the condition (\ref{condition1}) reduce to surface terms
\begin{align}
&\int\! dV_{c} |\textbf{E}|^{2}=\! \int \! dS_{1} \Phi^{*} E_{n}^{c},
\nonumber\\
&\int\! dV_{s} |\textbf{E}|^{2}=\! \int \! dS_{2} \Phi^{*} E_{n}^{s} - \! \int \! dS_{1} \Phi^{*} E_{n}^{s},
\end{align}
where $E_{n}^{j}(S_{i})=-\nabla_{jn}\Phi(S_{i})$ is the field outward normal component at the $i$th interface in the $j$th medium side. Note that $E_{n}^{s}(S_{1})=(\varepsilon_{c}/\varepsilon_{s})E_{n}^{c}(S_{1})$ and $E_{n}^{s}(S_{2})=(\varepsilon_{d}/\varepsilon_{s})E_{n}^{d}(S_{2})$.
The  ratio of integrated field intensities in the shell and core regions takes the form
\begin{equation}
L=\frac{\!\int\! dV_{s} |\textbf{E}|^{2}}{\!\int\! dV_{c} |\textbf{E}|^{2}}
=
\frac{\varepsilon_{d}}{\varepsilon_{s}}\frac{\int \! dS_{2} \Phi^{*} E_{n}^{d}}{\int \! dS_{1} \Phi^{*} E_{n}^{c}} - \frac{\varepsilon_{c}}{\varepsilon_{s}},
\end{equation}
where the potentials $\Phi$ are continuous at the interfaces. For nanostructures whose shape permits separation of variables, the potential  can be written as $\Phi (\bm{r})=R (\xi)\Sigma(\eta,\zeta)$, where $\xi$ is the radial (normal) coordinate and the pair $(\eta,\zeta)$ parametrizes the surface. With surface area element $dS=h_{\eta}h_{\zeta} d\eta d\zeta$ and normal derivative $\nabla_{n}=h_{\xi}^{-1}(\partial/\partial\xi)$, where $h_{i}$ are the scale factors ($i=\xi,\eta,\zeta$), the fraction of integrals   takes the form
\begin{align}
I=\frac{\int \! dS_{2} \Phi^{*} E_{n}^{d}}{\int \! dS_{1} \Phi^{*} E_{n}^{c}}
=
&\frac{R_{d} (\xi_{2})R_{d}' (\xi_{2})}{R_{c} (\xi_{1})R_{c}' (\xi_{1})}
\nonumber\\
\times
&\frac{\int \! \int \! d\eta_{2} d\zeta_{2} (h_{\eta_{2}}h_{\zeta_{2}}/h_{\xi_{2}})  |\Sigma |^{2}}
{\int \! \int \! d\eta_{1} d\zeta_{1} (h_{\eta_{1}}h_{\zeta_{1}}/h_{\xi_{1}})  |\Sigma |^{2}}.
\end{align}
Below we outline evaluation of $L$ for core-shell NP  described by two confocal prolate spheroids with semi-axises  $a$   and  $b$. The two shell surfaces corresponds to $\xi_{1}=a /f $ and  $\xi_{2}=sa /f $, where $f =\sqrt{a^{2}-b ^{2}}$ is half distance between the foci, and $s>1$ characterizes the shell thickness. Evaluation of  angular integrals yields
\begin{equation}
I
= \frac{R_{d} (\xi_{2})R_{d}' (\xi_{2})}{R_{c} (\xi_{1})R_{c}' (\xi_{1})} \,
\frac{\xi_{2}^{2}-1}{\xi_{1}^{2}-1}.
\end{equation}
For longitudinal dipole mode ($l=1$, $m=0$), we have $R_{c} = P_{1}(\xi)$ for $\xi<\xi_{1}$, $R_{s} = AP_{1}(\xi)+BQ_{1}(\xi)$ for $\xi_{2}<\xi<\xi_{2}$, and $R_{d} = CQ_{1}(\xi)$ for $\xi>\xi_{2}$, yielding
\begin{equation}
L=C^{2}\, \frac{\varepsilon_{d}}{\varepsilon_{s}}\,\frac{Q_{1}(\xi_{2})Q'_{1}(\xi_{2})}{P_{1}(\xi_{1})P'_{1}(\xi_{1})} \, \frac{\xi_{2}^{2}-1}{\xi_{1}^{2}-1} - \frac{\varepsilon_{c}}{\varepsilon_{s}},
\end{equation}
where $C$ and $\varepsilon_{c}(\omega_{pl})$ are determined from the continuity of $R_{i}$ and $\varepsilon_{i}R'_{i}$ across the interfaces.



\end{document}